\documentclass[%
 reprint,
 amsmath,amssymb,
 aps,
 pra,
]{revtex4-2}

\usepackage{graphicx}
\usepackage{dcolumn}
\usepackage{bm}
\usepackage{hyperref}

\newcommand{\zh}{\bm}

\newcommand{\mee}{{E}}

\newcommand{\dee}{{\varepsilon}}

\newcommand{\zhr}{{\zh r}}

\newcommand{\zhe}{{\zh e}}

\newcommand{\zhp}{{\zh p}}
\newcommand{\zhk}{{\zh k}}

\newcommand{\zhA}{{\zh A}}

\newcommand{\Br}[1]{(\ref{#1})}
\newcommand{\Eq}[1]{Eq.\ (\ref{#1})}

\newcommand{\Fig}[1]{Fig.\ \ref{#1}}
\newcommand{\txt}[1]{{\rm #1}}

\begin{document}

\title{Investigation of two-photon electron capture by H-like uranium}

\author{ Konstantin\ N.\ Lyashchenko $^{1,2,3}$}
\author{Oleg\ Yu.\ Andreev $^{3,2}$}
\author{Deyang\ Yu $^{1,4}$}

\affiliation{$^{1}$Institute of Modern Physics, Chinese Academy of Sciences, Lanzhou 730000, China}
\affiliation{${}^2$Petersburg Nuclear Physics Institute named by B.P. Konstantinov of National Research Centre ``Kurchatov Institute'', mkr. Orlova roshcha 1, Gatchina, 188300, Leningrad District, Russia}
\affiliation{$^{3}$St.\ Petersburg State University, 7/9 Universitetskaya nab., St. Petersburg, 199034, Russia}
\affiliation{$^{4}$University of Chinese Academy of Sciences, Beijing 100049, China}

\date{\today}

\begin{abstract}
We present a study of two-photon electron capture by H-like uranium ions.
The energy of the incident electron was chosen to be in the region with the most significant contribution of the dielectric recombination. 
We studied the photon emission spectrum, including the main resonance groups corresponding to the cascade transition, and the low-energy photon region, where the infrared divergence required special processing.
The calculations were performed within the framework of QED theory.
The importance of generalized Breit interaction was discussed.
We investigated the roles of the dielectric recombination and the radiative recombination.
We introduced and investigated the resonance approximation and the single-photon approximation, which are commonly used to describe radiation spectra.
\end{abstract}


\maketitle

\section{Introduction}
The radiative electron transitions in ions and atoms are the fundamental processes in atomic physics. 
The one- and two-photon transitions determine the majority of possible radiative transitions.
The simplest system for observing such transitions is H-like ions or a hydrogen atom. 
The one-photon transitions are direct transitions which occur without formation of any intermediate state, while the two- and more photon transitions can proceed both through the formation (cascade transitions) and without the formation (noncascade transitions) of intermediate states.

The radiative transitions include processes in which an incident electron emits one or more photons passing into a bound state (radiative electron capture) \cite{eichler07} or into a continuum state with lower energy (bremsstrahlung) \cite{sommerfeld1931,haug2004book}.
The first theoretical description of the two-photon mechanism was presented in \cite{goppertmayer31,goppertmayer31_2}.
Consideration of systems with two or more electrons leads to the appearance of such types of transitions in which the interelectron interaction can play a significant role.
The radiative transitions in such systems can also be distinguished by the number of emitted photons and by the formation of intermediate states.
The role of the interelectron interaction can be very different.
In this work we focus on radiative transitions in which the interelectron interaction plays a crucial role.

We consider the two-photon electron capture by H-like ions.
It is customary to distinguish two channels in this process: radiative recombination (RR) and dielectric recombination (DR).
The RR is a nonresonant channel of the electron capture in which electron is directly captured into a bound state of the ion
\begin{eqnarray}
e^{-}(\varepsilon)+\txt{U}^{91+}(1s)
&\to&\label{eqn23ef}
\txt{U}^{(90+)}(1s,\, nl) +\gamma 
\,,
\end{eqnarray}
where $e^{-}$ denotes the incident electron with the energy $\varepsilon$, $\gamma$ is the emitted photon and $n$, $l$ correspond to the principal quantum number and the orbital momentum of the one-electron state, respectively.
In contrast, the DR is a resonant channel in which the electron capture proceeds through the formation of a doubly excited state
\begin{eqnarray}
e^{-}(\varepsilon)+\txt{U}^{90+}(1s)
&\to&\nonumber
\txt{U}^{90+}(nl,n'l')\\
&&\nonumber
\hspace{30pt}\downarrow
\\
&&\label{eqnwefwef}
\txt{U}^{90+}(1s,\, nl) +\gamma 
\,,
\end{eqnarray}
where $n,n'\ge 2$.
The DR can make a significant contribution to the cross section if the energy of the initial state is close to the energy of some of the doubly excited states.
We note that the division of the process into these two channels is quite conditional.
The interference between these channels can also be of importance.

The radiative electron capture in which only the RR channel makes significant contribution has been extensively investigated for many atomic systems \cite{eichler07,kroger2020PhysRevA.102.042825,Maiorova2023PhysRevA.107.042814}.
In systems with two or more electrons, the DR channel acquires special importance and attracts considerable attention from experimenters \cite{Nakamura2008PhysRevLett.100.073203,bernhardt11,Mahmood_2012,Lindroth2020PhysRevA.101.062706,hu2022PhysRevA.105.L030801,Nakamura2023PhysRevLett.130.113001}.
It also plays a significant role for the description of the laboratory plasma and that which is observed in astrophysics \cite{hitomi2018}.
The first measurements of the DR cross section were reported in \cite{Nakamura2008PhysRevLett.100.073203,bernhardt11}.
In these works, the theoretical predictions about the large contribution of the Breit interaction and the interference between the resonances were convincingly confirmed experimentally.
Recently, the energy spectra of the emitted photon in DR have become the object of experimental research.
In the work \cite{Nakamura2023PhysRevLett.130.113001}, it was reported about the measurement of photon emission spectra for the DR with Be-like lead ions.

The DR was also extensively studied theoretically.
In particular, the calculation of DR cross sections for the highly charged ions were presented in
\cite{chen90,pindzola90,zimmerer90,karasiev92,zakowicz04,andreev09p042514,bernhardt11}.
In the majority of these works, the process of electron capture was considered within the single-photon approximation, i.e., only the one-photon transitions to the singly excited states were taken into account, while the further decay of these states was ignored \Br{eqnwefwef}.
Hence, only the emission of the resonant photon was investigated.
Going beyond this approximation was discussed in \cite{zakowicz04}, where the process was analyzed within the framework of the resonance approximation (with disregard for the noncascade transitions and their interference with the main channel).
In the most accurate experiments on the DR \cite{Nakamura2008PhysRevLett.100.073203,bernhardt11}, the cross section was measured by recording the change in ion charge after the electron capture, while the photon emission spectrum was not recorded.
The single-photon approximation quite precisely describes these experimental data.
However, it is not appropriate for the investigation of the photon emission spectrum.
In this paper, we study for the first time the photon emission spectrum of the electron capture by H-like ions, without using such common approximations as the single-photon approximation and the resonance approximation.
We investigate these approximations, comparing their results with the results of the full calculation, and discuss the conditions of their applicability.

The two-photon transitions have many additional properties compared to the one-photon transitions, which leads to their much wider participation in the atomic processes. Various properties of two-photon transitions are an actual subject of the experimental and theoretical research for both light and heavy ions and atoms \cite{goppertmayer31,Breit1940,Grynberg_1977,Nussbaumer1984,dunford89,Bottorff_2006,Hirata2008PhysRevD.78.023001,Zalialiutdinov2014PhysRevA.89.052502,andreev08pr,Martin2018PhysRevApplied.9.014019,Sommerfeldt2020PhysRevA.102.042811,hitomi2018,Kulkarni_2022}.
In this paper, we present an {\it ab initio} QED study of the two-photon electron capture in the presence of the DR channel using the example of uranium ions.
In particular, we show that the DR channel leads to quantitative and serious qualitative changes in the total and differential cross section.

The paper is organized as following.
In the next section we present the QED approach which was used for the calculation.
The third section is divided into subsections, where we present and discuss our results.
First, we show the total cross section of the two-photon electron capture as a function of the energy of the incident electron and discuss the electron energy selected for further study.
The resonance structure of the differential cross section as a function of the emitted photon energy is investigated.
Then we discuss the two-photon emission cross section for the low-energy photons, since the infrared divergence that occurs in this energy region requires a special approach.
Further we discuss the contribution of the DR channel to the differential cross section comparing it with the RR channel.
Next, we investigate the role of the Breit interaction for the emission spectrum.
In the last two subsections, we introduce and study the resonance and single-photon approximation, commonly used to describe the radiation spectrum and total cross section, respectively.
In conclusion, we provide a brief summary.

\section{Theory}
We present an {\it ab initio} QED study of the two-photon electron capture.
We performed the calculation of the differential cross section depending on the energy of the emitted photon.
The considered process is schematically described as
\begin{eqnarray}
e^{-}(\varepsilon)+\txt{U}^{91+}(1s)
&\to&\label{eqn170705n01x}
\cdot\cdot\cdot
\,\to\,
\txt{U}^{90+}(1s)^2 + \gamma+\gamma'
\,,
\end{eqnarray}
where the initial state is the incident electron with the energy ($\dee$) and the bound $1s$-electron.
The final state is the two-electron ion in the ground $(1s)^2$ state.
A special attention is paid to the energy region of the emitted photon where the electron capture proceeds through the formation of one of the singly excited states.
In this case, the process can be described as
\begin{eqnarray}
e^{-}(\varepsilon)+\txt{U}^{91+}(1s)
&\to&\nonumber
\txt{U}^{90+}(1s,\,nl) + \gamma \\
&&\nonumber
\hspace{30pt}\downarrow
\\
&&\label{eqn170705n01}
\txt{U}^{90+}(1s)^2 + \gamma+\gamma'
\,.
\end{eqnarray}
In the DR channel, this process proceeds through the additional formation of a doubly excited state
\begin{eqnarray}
&&\nonumber
e^{-}(\varepsilon)+\txt{U}^{91+}(1s)
\,\to\,
\txt{U}^{90+}(nl,n'l')\,\to
\\
&\to&\label{eqn170705n00}
\txt{U}^{90+}(1s,n'l') + \gamma
\,\to\,
\txt{U}^{90+}(1s)^2 + \gamma+\gamma'
\,,
\end{eqnarray}
where $n,n'\ge 2$. 
The process of the two-photon electron capture \Br{eqn170705n01x} including its subprocesses given by \Br{eqn170705n01} and \Br{eqn170705n00} is treated uniformly as a composite process.

The relativistic units are used throughout the paper unless otherwise stated.

\begin{figure}[h]
\includegraphics[width=15pc]{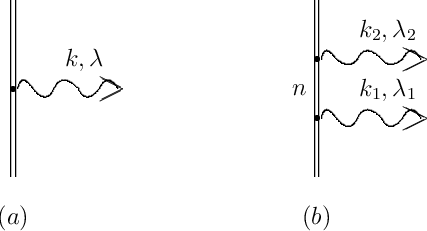}
\caption{The Feynman graphs corresponding to the one-photon (a) and two-photon (b) emission.
The double line represents the electron in the field of the atomic nucleus.
The wavy lines with arrows correspond to the emitted photons with the momentum 4-vector ($k$) and polarization ($\lambda$).
The letter $n$ indicates the summation over the complete Dirac spectrum (the electron propagator).
}
\label{fig_1el_2ph}
\end{figure}   

\begin{figure}[h]
\includegraphics[width=15pc]{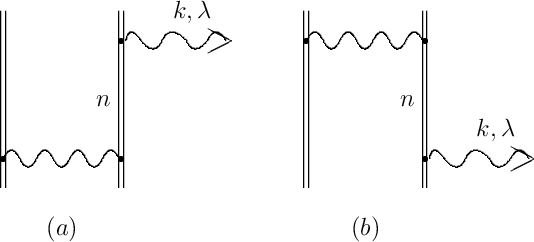}
\caption{The Feynman graphs corresponding to the one-photon emission in He-like ions.
The wavy line between the vertexes gives the photon propagator. The other notations are the same as in
\Fig{fig_1el_2ph}.
}
\label{fig_2el_1ph}
\end{figure}   

\begin{figure}[h]
\includegraphics[width=20pc]{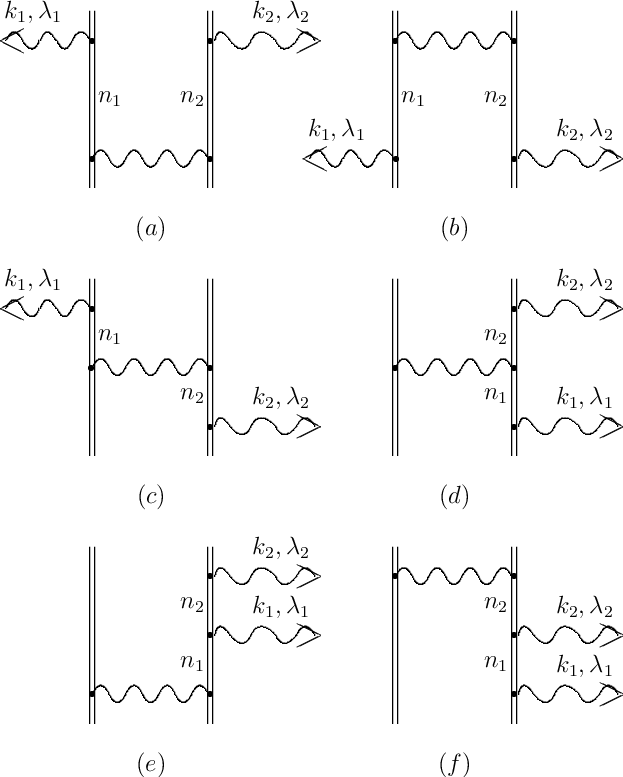}
\caption{The Feynman graphs describing the two-photon emission with the photon exchange. The notations are the same as in
\Fig{fig_2el_1ph}.
}
\label{fig_2el_2ph}
\end{figure}   
 
The cross section of the two-photon electron capture was calculated with the use of the line-profile approach (LPA).
A detailed description of LPA is presented in
\cite{andreev08pr}.
In this work, we generalize this method for the two-photon electron capture and present the main points of this approach.
The one- and two-photon emissions in H-like ions are given by the Feynman graphs depicted in
\Fig{fig_1el_2ph}.
The Feynman graphs describing the two-photon emission with one-photon exchange correction are presented in
Figs.~\ref{fig_2el_1ph} and \ref{fig_2el_2ph}.

The Furry picture \cite{furry51} was used in which the interaction of electrons with the electric field of the atomic nucleus was fully taken into account.
The two-electron wave functions of the final $(1s)^2$ state and all the intermediate states in the zero-order perturbation theory are expressed in the $j$--$j$ coupling scheme 
\begin{eqnarray}
\Psi^{(0)}_{JMn_1j_1l_1n_2j_2l_2}(\zhr_1,\zhr_2)
&=&\nonumber
N\sum\limits_{m_1m_2}
C^{JM}_{j_1m_1j_2m_2}\\
&&\label{jjcs}
\hspace{-60pt}
\times
\det\{\psi_{n_1j_1l_1m_1}(\zhr_1),\psi_{n_2j_2l_2m_2}(\zhr_2)\}
\,,
\end{eqnarray}
where the one-electron wave functions $\psi_{njlm}$ are the solutions of the Dirac equation,
$n$, $j$, $m$ denote the principal quantum number (or the energy, in the case of continuum electrons),
the total angular momentum and its projection, respectively.
Further, $J$ and $M$ are the total angular momentum of the two-electron configuration
and its projection,
$N$ is the normalizing constant, which is equal to $1/\sqrt{2}$ for non equivalent electrons and to $1/2$ for equivalent electrons.
The symbols $C^{JM}_{j_1m_1j_2m_2}$ denote the Clebsch-Gordan coefficients \cite{Varshalovich1988QuantumTO}.

The initial state of the electron system ($1s$ and $e^{-}(\varepsilon)$) includes the incident electron with the certain momentum ${\zhp}$ and polarization $\mu$ (in the asymptotic $r\to\infty$). Its wave function can be written as
\begin{eqnarray}
\Psi^{(0)}_{njlm, {\zhp} \mu}(\zhr_1,\zhr_2)
&=&
\frac{1}{\sqrt{2}} \det\{ \psi_{njlm}(\zhr_1), \psi_{{\zhp} \mu}(\zhr_2) \}
\,,
\end{eqnarray}
where $\psi_{njlm}$ is the wave function of the bound electron, $\psi_{{\zhp} \mu}$ is the wave function of the continuum electron.
It is convenient to represent this wave function as an expansion over the wave functions in the $j$--$j$ coupling scheme
\begin{eqnarray}
\Psi^{(0)}_{njlm, {\zhp} \mu}(\zhr_1,\zhr_2)
&=&\nonumber
\int d\epsilon
\sum\limits_{JMj'l'm'}
C^{JM}_{jmj'm'}
\\
&\times&
a_{{\zhp} \mu,\epsilon j'l'm'}
\Psi^{(0)}_{JMnjl\epsilon j' l'}(\zhr_1,\zhr_2)
\,,
\end{eqnarray}
where the coefficients $a_{{\zhp} \mu,\epsilon j'l'm'}$ read as
\begin{eqnarray}
a_{{\zhp} \mu,\epsilon j'l'm'}
&=&\nonumber
\frac{(2\pi)^{3/2}}{\sqrt{p\varepsilon}} i^{l'} e^{i\phi_{j'l'}} 
\left( \Omega^{+}_{j'l'm'}({\zhp}) \upsilon_{\mu} ({\zhp}) \right)\\
&\times&
\delta(\epsilon-\varepsilon)
\,,
\end{eqnarray}
$\Omega^{+}_{j'l'm'}({\zhp})$ is the spherical spinor, $\upsilon_{\mu} ({\zhp})$ is the spinor
with projection ($\mu$) on the direction of the electron momentum ${\zhp}$,
the phases $\phi_{j'l'}$ are the Coulomb phases, and $\varepsilon$ is the energy of the incident electron \cite{akhiezer65b}. 

The interelectron interaction plays an essential role for the formation of doubly excited (autoionizing) states.
Therefore, it should be taken into account accurately.
The doubly excited states are usually quasidegenerate.
Accordingly, the quasidegenerate QED perturbation theory should be used for the description of the DR process.
Applying the quasidegenerate perturbation theory within the LPA, we introduce a set of two-electron configurations (the set $g$) which includes the reference state (the initial state, final state and some of the intermediate states corresponding to the considered resonances) and all the two-electron configurations with the energies close to the reference states.
For taking into account the interaction of the reference states with the quantized fields, the matrix $V$ is introduced which is determined by the one and two-photon exchange, electron self-energy and vacuum polarization matrix elements and other QED corrections.
The matrix $V$ is derived order by order within the framework of the QED perturbation theory \cite{andreev08pr}. 
It is convenient to present the matrix $V$ as a block matrix
\begin{eqnarray}
V 
&=& \nonumber
\left(
 \begin{array}{cc}
  V_{11}  & V_{12}  \\
  V_{21}  & V_{22}  \\
 \end{array}
\right)\\
&=&
\left(
 \begin{array}{cc}
  V^{(0)}_{11}+ \Delta V_{11}  & \Delta V_{12}  \\
  \Delta V_{21}  & V^{(0)}_{22}+ \Delta V_{22}  \\
 \end{array}
\right)
\,,
\end{eqnarray}
where matrix $V_{11}$ is defined on the set $g$.
The matrix $V_{11}$ is a finite matrix and can be diagonalized numerically
\begin{eqnarray}
V^{\txt{diag}}_{11}
&=&
B^T V_{11} B, \,\,\, B^T B = I.
\end{eqnarray}
Then, the standard perturbation theory can be applied for the diagonalization of the infinite matrix $V$. The reference states are described by the corresponding eigenvectors of this matrix
\cite{andreev09p042514}
\begin{eqnarray}
\Phi_{ n_g}
&=&\nonumber
\sum\limits_{ k_g \in g}
B_{k_g {n_g}}\Psi^{(0)}_{ k_g}\\
&&\label{eigenphi}
+
\sum\limits_{ k\notin g, { l_g\in g}}
\!\!
[\Delta V]_{ k { l_g}}
\frac{B_{{l_g} { n_g}}}
{{\mathcal{E}}_{n_g}-E^{(0)}_{k}}
\Psi^{(0)}_{k}
+
\cdots
\,,
\end{eqnarray}
where $n_g\equiv(JMj_1j_2l_1l_2n_1n_2)$ is a complex index representing the complete set of quantum numbers describing the reference state $n_g$,
the indices $k$, $l_g$ describe the two-electron configurations:
the index $l_g$ runs over all configuration of the set $g$;
the index $k$ runs over all configuration not included in the set $g$
(this implies the integration over the positive- and negative energy continuum). 
Here, $E^{(0)}_k$ is the energy of the two-electron configuration
in the zeroth order:
sum of the one-electron Dirac energies. 
The eigenvectors $\Phi_{n_g}$ corresponding to the reference states are used for calculation of the amplitude of the process under consideration.
The eigenvalues ${\mathcal{E}}_{n_g}=E_{n_g}-\frac{i}{2}\Gamma_{n_g}$ are complex, where $E_{n_g}$ is interpreted as the energy of the corresponding energy level (whith the interelectron interaction and QED corrections are taken into account) and $\Gamma_{n_g}$ is the width of the energy level. The calculations of the energy levels for the He-like ions were presented in works \cite{lindgren95,andreev04,andreev09p042514,artemyev05,lyashchenko2021PhysRevA.104.012818}.

Within the LPA the amplitude of the two-photon electron capture is expressed as \cite{andreev08pr}
\begin{eqnarray}
U_{FI}
&=&\nonumber
\sum_{N}\frac{\left(A^{(k_1,\lambda_1)*}\right)_{FN} \left(A^{(k_2,\lambda_2)*}\right)_{NI} }{E_{F}+\omega_1-E_{N}+\frac{i}{2}\Gamma_{N}}
\\
&+&\label{3}
\sum_{N}\frac{\left(A^{(k_2,\lambda_2)*}\right)_{FN} \left(A^{(k_1,\lambda_1)*}\right)_{NI} }{E_{F}+\omega_2-E_{N}+\frac{i}{2}\Gamma_{N}}
\,,
\end{eqnarray}
where $\mee_F$ is the energy of the final state of the electron system and the two-electron matrix element of the photon emission $\left(A^{(k,\lambda)*}\right)_{UD}$ reads as
\begin{eqnarray}
\left(A^{(k,\lambda)*}\right)_{UD}
&=&\nonumber
e\int d^3 {\bf r}_1 d^3 {\bf r}_2 \,
\overline{\Phi}_{U} ({\bf r}_1, {\bf r}_2)  
\\
&&\nonumber
\hspace{-40pt}
\times
\left(
\gamma^{(1)\nu}A^{(k,\lambda)*}_{\nu}(\zhr_1)
+
\gamma^{(2)\nu}A^{(k,\lambda)*}_{\nu}(\zhr_2)
\right)
\\
&&\label{Eq576789}
\hspace{-40pt}
\times
\Phi_{D} ({\bf r}_1, {\bf r}_2)
\,.
\end{eqnarray}
In \Eq{Eq576789} $\gamma^{(i)\nu}$ are the Dirac $\gamma$-matrices acting on the one-electron wave function of the argument $\zhr_i$.
The photon wave function
$A^{(k,\lambda)\nu}=(A_0^{(k,\lambda)},{\zhA}^{(k,\lambda)})$
in the transverse gauge reads as
\begin{eqnarray}
A_0^{(k,\lambda)}(\zhr)
&=&
0
\,,\,\,\,
{\zhA}^{(k,\lambda)}({\zhr})
\,=\,
\sqrt{\frac{2\pi}{\omega}} e^{i \zhk\zhr} \zhe^{(\lambda)}
\,,
\end{eqnarray}
where $\zhk$ is the photon wave vector, $\omega=|\zhk|$ is the photon energy (frequency),
$\zhe^{(\lambda)}$ is the polarization vector.

The summations in \Eq{3} run over the complete basis set constructed from the two-electron functions \Eq{eigenphi}.
However, in this work it is sufficient to take into account only the two-electron $(n_1 l_1, \, n_2 l_2)$ states where the principal quantum number $n_1$ of the first electron is equal $1$ and $2$.
The quantum numbers of the second electron run over the complete Dirac spectrum.

The fully differential cross section is connected with the amplitude as
\begin{eqnarray}
\frac{d^4\sigma}{d\omega_1 d\Omega_1 d\omega_2 d\Omega_2}
&=&\nonumber
\delta(\omega_1+\omega_2-E_{I}+E_{F})\\
&\times&\label{2}
\left|U_{FI}\right|^2
\frac{\varepsilon}{p}
\frac{\omega_1^2 \omega_2^2}{(2\pi)^5}
\,,
\end{eqnarray}
where $E_{I}$, $E_{F}$ are the energies of the initial and final state of the electron system, respectively,
$\Omega_{1,2}$ are the solid angles of the emitted photons.

Accordingly, the differential cross section of the two-photon electron capture is given by
\begin{eqnarray}
\frac{d\sigma}{d\omega_1}
&=&\nonumber
\frac{1}{4}\sum_{\mu,m_b}\sum_{\lambda_1,\lambda_1}
\int d\omega_2 d\Omega_1 d\Omega_2\\
&\times&\label{Eq782267}
\frac{d^4\sigma}{d\omega_1 d\Omega_1 d\omega_2 d\Omega_2} 
\,.
\end{eqnarray}
The energies of the emitted photons are limited by the interval determined by the energy conservation law
\begin{eqnarray}
\omega_1+\omega_2
&=&\label{eqn231124n01}
E_{I}-E_{F}
\,.
\end{eqnarray}
The photon energy spectrum is continuous and limited by the energy range $[0,\omega_{\txt{max}}]$, where
$\omega_{\txt{max}}=E_{I}-E_{F}$.
We note that if one of the photon is registered, then the energy of the second one is determined by
\Eq{eqn231124n01}.
Since the two-photon states $(\zhk_1,\lambda_1;\zhk_2,\lambda_2)$ and $(\zhk_2,\lambda_2;\zhk_1,\lambda_1)$ are identical, the differential cross section \Eq{Eq782267} is symmetric with respect to the center of the energy interval $[0,\omega_{\txt{max}}]$.

The summation over the complete Dirac spectrum was performed using a finite basis set for the
Dirac equation constructed from B-splines \cite{johnson88p307,shabaev04}.
The implementation of this method for the two-electron system is described in our previous work  \cite{lyashchenko2021}.
The real part of the electron self-energy correction were taken into account with the use of \cite{Shabaev2013PhysRevA.88.012513}.

\section{Results and discussion}
We theoretically study the two-photon electron capture by H-like highly charged ions using the uranium ions as an example.
We have calculated the differential cross section with respect to the energies of the emitted photons.
Particular attention was paid to the collision energies at which the DR makes a significant contribution to the cross section.

\subsection{Dielectronic recombination}
\begin{figure}[h]
\includegraphics[width=20pc]{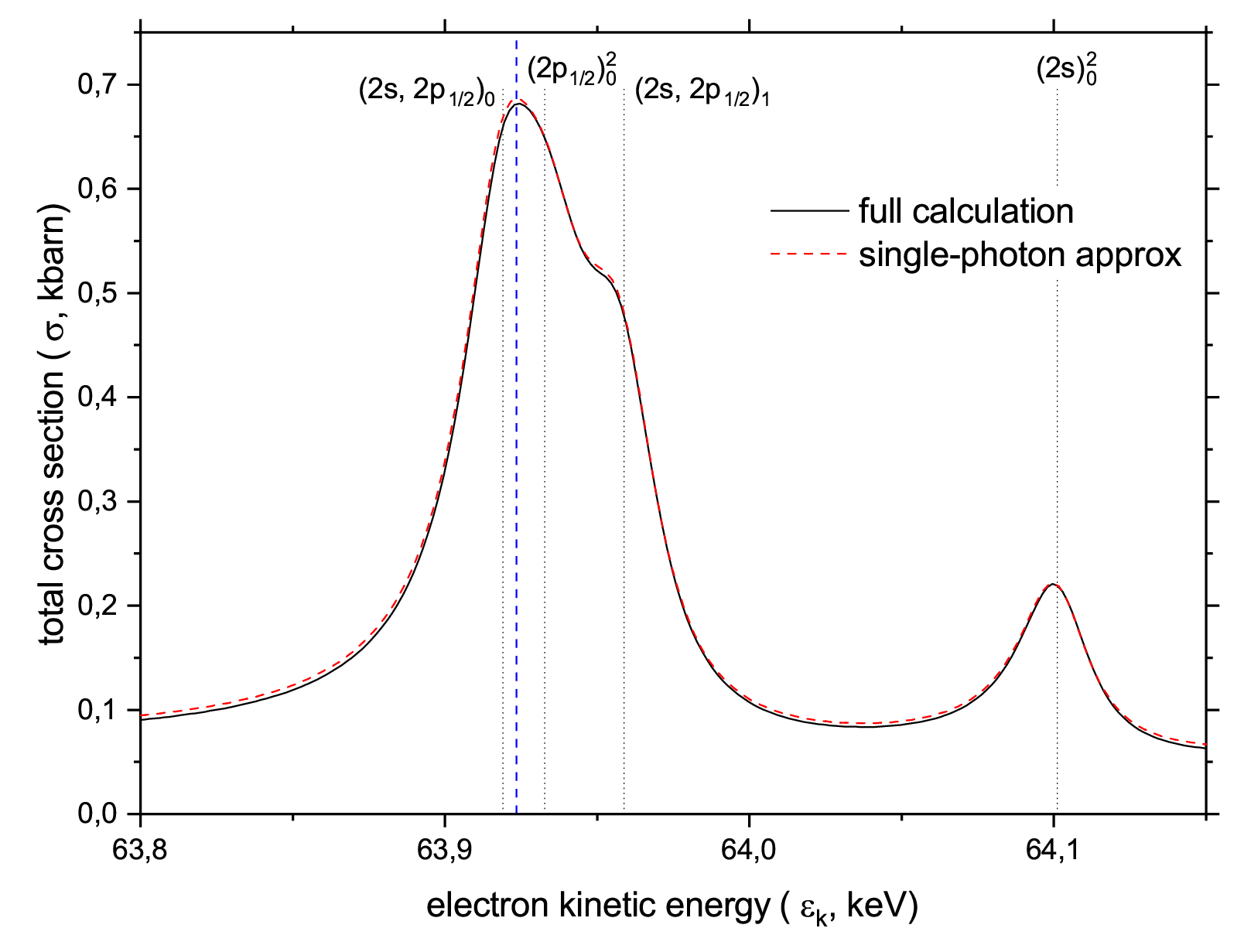}
\caption{Total cross section (in kbarn) of the electron capture as a function of the kinetic energy of the incident electron for uranium ion. The solid (black) curve corresponds to the total cross section of the two-photon electron capture to the ground state. The dashed (red) curve represents the total cross section calculated within the single-photon approximation (only the capture to the single excited state with total angular momentum $J\ne 0$ are taken into account, see Section \ref{single-photon_approximation}). The dotted (black) vertical lines show the positions of the DR resonances corresponding to the doubly excited states. The dashed (blue) vertical line indicates the kinetic energy of the incident electron, chosen for the study of the differential cross section of two-photon electron capture.}
\label{fig00}
\end{figure}

The DR is a resonant channel of the electron capture which proceeds through the formation of a doubly excited state.
The corresponding total cross section shows resonances at the incident electron energies, at which the energy of the initial state approaches the energy of one of the doubly excited states.
In Fig.~\ref{fig00} we present the total cross section (the solid (black) curve) of electron capture for the DR energy region for the uranium ions.
The resonances of the cross section demonstrate the contributions of the particular doubly excited states which are indicated by the vertical dotted (black) lines.
For a detailed study of the photon emission spectrum and the effect of the DR, the collision energy was selected in such a way that it corresponded to the strongest resonance of the cross section.
This energy is indicated by the vertical dashed (blue) line in Fig.~\ref{fig00} and is equal to $63.924\,$keV.
The dashed (red) curve represents the results obtained within the framework of the single-photon approximation, which is discussed in the subsection~\ref{single-photon_approximation}.

\begin{figure*}[ht]
\includegraphics[width=40pc]{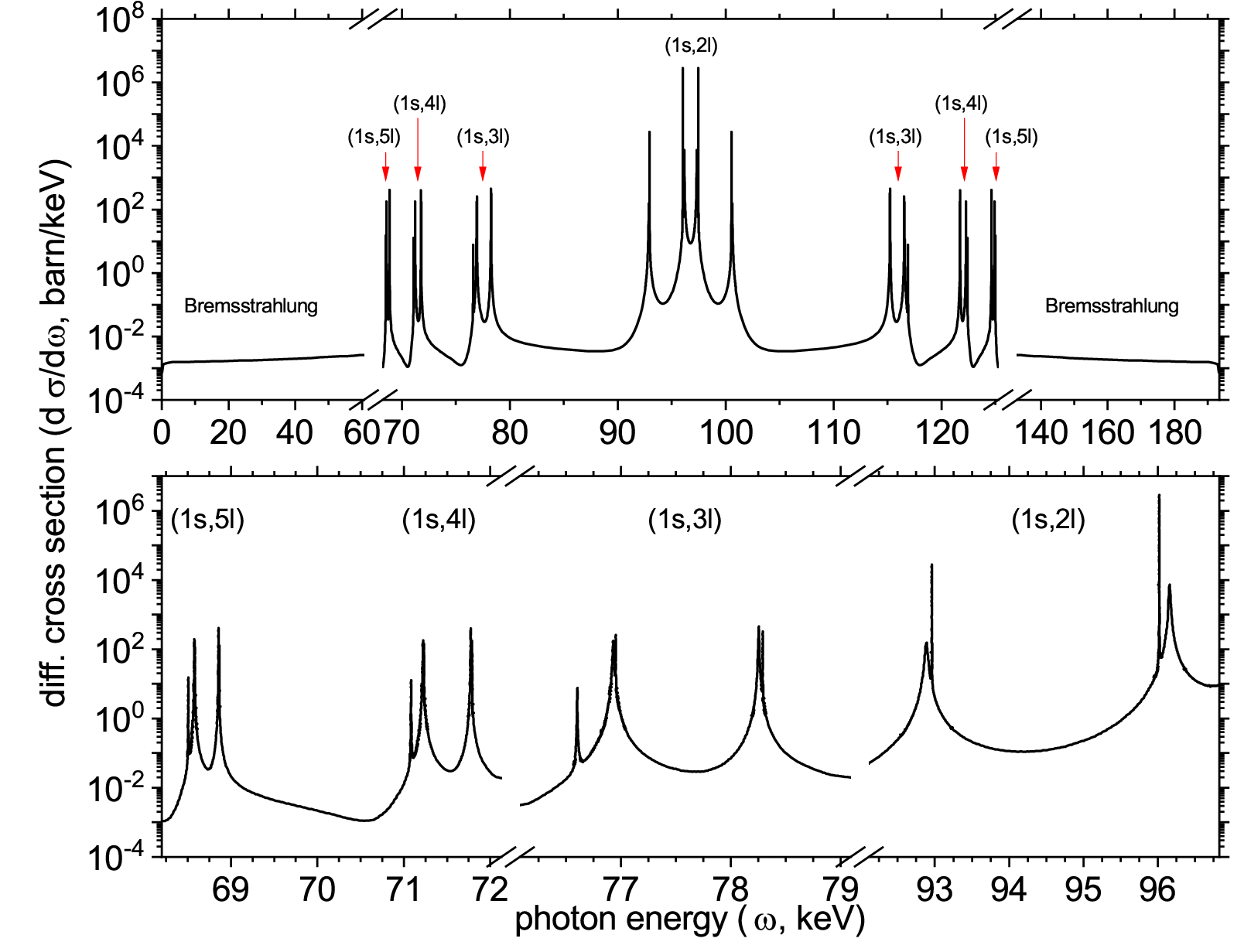}
\caption{Differential cross section (in barn/keV) of the two-photon electron capture as a function of the emitted photon energy for uranium ion. The electron kinetic energy is $\varepsilon_{\txt{k}}=63.924\,$keV, which is close to the DR resonances corresponding to the $(2p_{1/2})^2$, $(2s 2p_{1/2})_0$, and $(2s 2p_{1/2})_1$ states.}
\label{fig1}
\end{figure*}

In Fig.~\ref{fig1}, we present the two-photon emission differential cross section with respect to the energy of the emitted photon (the energy of the paired photon is determined by \Eq{eqn231124n01}). Accordingly, the emitted photon energy is limited by the interval $[0,\omega_{\txt{max}}]$ and the differential cross section is symmetric with respect to the center of this interval.
The noticeable resonant structure indicates the various cascade transitions, i.e., the transition from the initial state to one of the singly excited $(1s\,nl_j)_J$ states:
\begin{eqnarray}
\txt{U}^{91+}(1s)+e^{-}(\varepsilon)
&\to&\nonumber
\txt{U}^{90+}(1s,\,nl_j)_J + \gamma\\
&&\nonumber
\hspace{30pt}\downarrow\\
&&
\txt{U}^{90+}(1s)^2 + \gamma+\gamma'
\,,
\end{eqnarray}
where $2 \leq n \leq 5$, $0 \leq l \leq 4$ and $J\geq 1$.
The intermediate states with $J=0$ do not contribute to the considered two-photon recombination (with the final $(1 s)^2$ state), since such states decay to the ground state with emission of two or more photons.
The cut-out energy regions on the upper panel of Fig. \ref{fig1} contain infinite number of resonances corresponding to the states with $n>5$.
When $n\to\infty$, the positions of the resonances ($\omega_n$) corresponding to the $(1s,nl_j)_J$ states approach the kinetic energy of the incident electron ($\omega_n\to\dee_{\txt{k}})$, in the left half of the energy interval, and tend to $\omega_{\txt{max}}-\dee_{\txt{k}}$ in the right half.
At the left and right edges of the energy interval, we can see a smooth behavior of the differential cross section.
These parts represent a radiative transitions to the lower energy continuum states (the bremsstrahlung) and the following radiative recombination to the ground state
\begin{eqnarray}
\txt{U}^{91+}(1s)+e^{-}(\dee)
&\to&\nonumber
\txt{U}^{91+}(1s)+e^{-}(\dee') +\gamma\\
&&\nonumber
\hspace{30pt}\downarrow\\
&&
\txt{U}^{90+}(1s)^2 + \gamma + \gamma'
\,,
\end{eqnarray}
where $\dee'<\dee$.
On the lower panel Fig. \ref{fig1} we present in detail four energy intervals with cascade resonances corresponding to the singly excited states with $n=2-5$.
The contribution of cascade transitions as well as the background value between the neighboring groups of peaks decrease with increase of the principal quantum number $n$.
The latter is mainly due to the interference between the groups of peaks.
This is discussed in subsection~\ref{contribution_dr}.

\subsection{Treatment of the infrared divergence}
\begin{figure}[h]
\includegraphics[width=20pc]{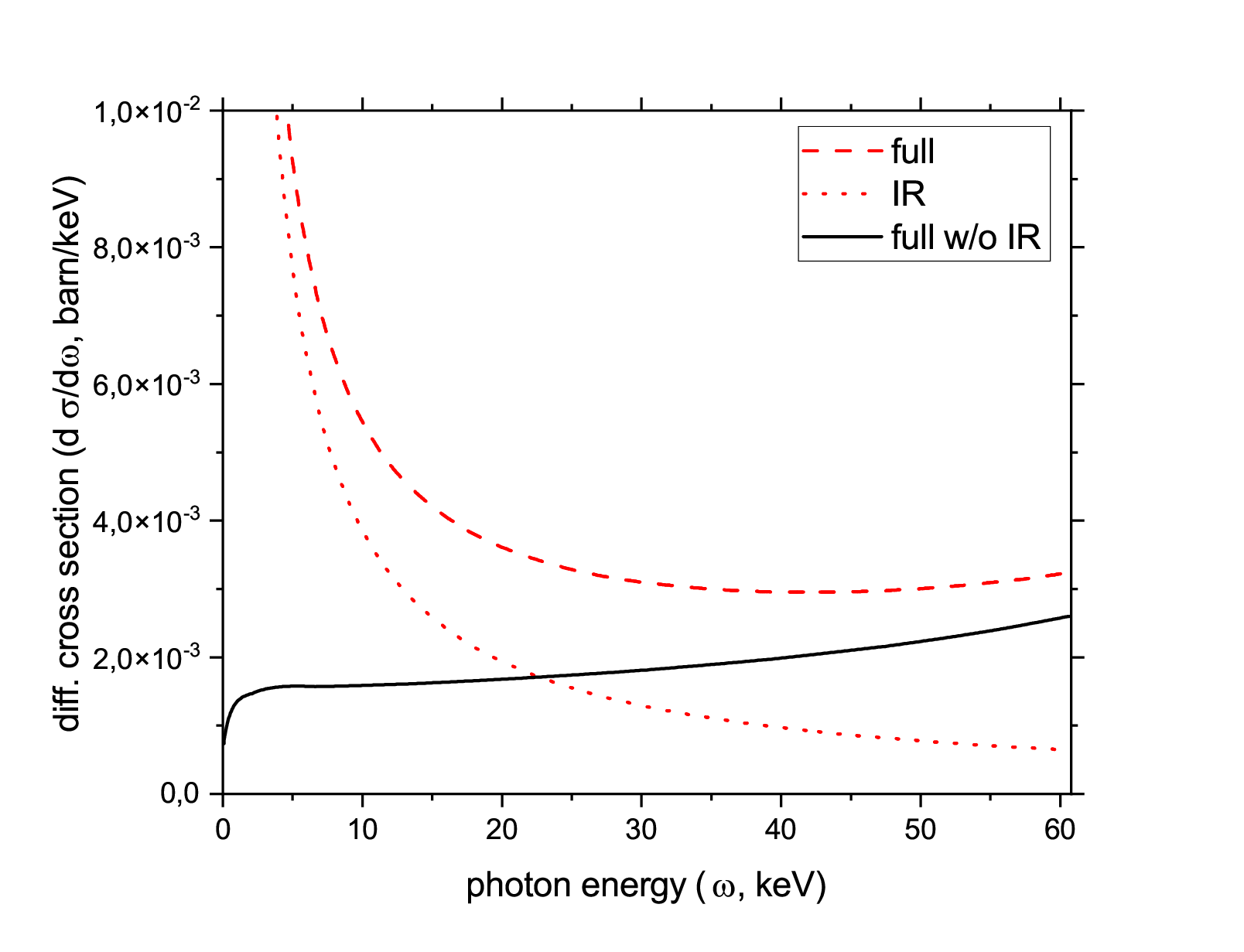}
\caption{The differential cross sections (in barn/keV) of the two-photon electron capture by H-like uranium ion with and without elimination of the infrared divergence. The dashed (red) curve corresponds to the differential cross section calculated with the amplitude in Eq.~(\ref{3}). The dotted (red) curve represents the divergence part of the dashed curve calculated by Eq.~(\ref{eq1293}). The solid (black) curve corresponds to the differential cross section after eliminating the infrared divergence (the difference between the dashed and dotted curves).}
\label{fig1_1}
\end{figure}

Special attention should be paid to the emission of  low-energy photons (soft photons), since the infrared divergence is manifested there \cite{bloch1937,jauch1976}. 
The reason for the infrared divergence lies in the use of the standard QED perturbation theory, which is based on the assumption that with an increase in the number of interactions between the quantized fields, the contributions of the corresponding terms decrease. This condition is not fulfilled when we consider the emission of soft photons \cite{jauch1976}.
Accordingly, the consideration of the energy region of the soft photon requires a reformulation of the QED perturbation theory.
The corresponding procedure is discussed in \cite{bloch1937,jauch1976,shabaev2000}.
In Fig. \ref{fig1_1} we explore the soft photon region for uranium ion in detail.
The dashed (red) curve, marked as `full', shows the cross section calculated within the framework of the standard QED perturbation theory. We can see that the cross section tends to infinity as the photon energy goes to zero. It can be deduced that the divergence part of the cross section has the following behavior
\begin{eqnarray}
\frac{d\sigma}{d\omega}
&\propto&
\frac{1}{\omega}
\,,\quad
\omega\to0
\,,
\end{eqnarray}
which leads to logarithmic divergence of the total cross section.
In one-electron case,
the reformulation of the perturbation theory can be reduced to the subtraction of the regularisation term, in which the wave function of a continuum electron (moving in the field of an atomic nucleus) is replaced by the wave function of free electron \cite{mandrykina2022}.
In the two-electron case, a similar procedure can be applied.
In the region of soft photon energies, the role of interelectron interaction for the initial and final states is relatively small (because cascade resonances are located far from this region).
If the interelectron interaction is negligible, then the cross section in two-electron case can be quite accurately reduced to a cross section for one-electron case with an additional factor $1/2$, which arises due to the different statistics for the initial and final states in these cases. Accordingly, following this procedure the regularization term reads as
\begin{eqnarray} 
\frac{d\sigma^{\txt{IR}}}{d\omega_1}
&=&\label{eq1293}
\frac{\alpha}{2\pi p \omega_1} 
\left(\varepsilon\ln{\frac{\varepsilon+p}{\varepsilon-p}}-2p\right)
\sigma^{\txt{1\gamma}} 
\,,
\end{eqnarray}
where $\sigma^{\txt{1\gamma}}$ denotes the total cross section of the one-photon radiative recombination to the ground state for one-electron case.
The cross section calculated with the replacement of the wave function of the continuum electron by the wave function of the corresponding free electron is represented in Fig. \ref{fig1_1} by a dotted (red) line, marked as `IR'.
The solid (black) curve, marked as `full w/o IR', gives the result of subtracting the regularization term from the cross section calculated within the framework of the standard QED perturbation theory.
The subtracted regularization term (after integration over the low-energy photons) can be canceled with the infrared divergent term arising from the radiation corrections to one-photon recombination of emitting electrons \cite{shabaev2000}.

\subsection{Contribution of the DR channel to the differential cross section}\label{contribution_dr}
\begin{figure}[ht]
\includegraphics[width=20pc]{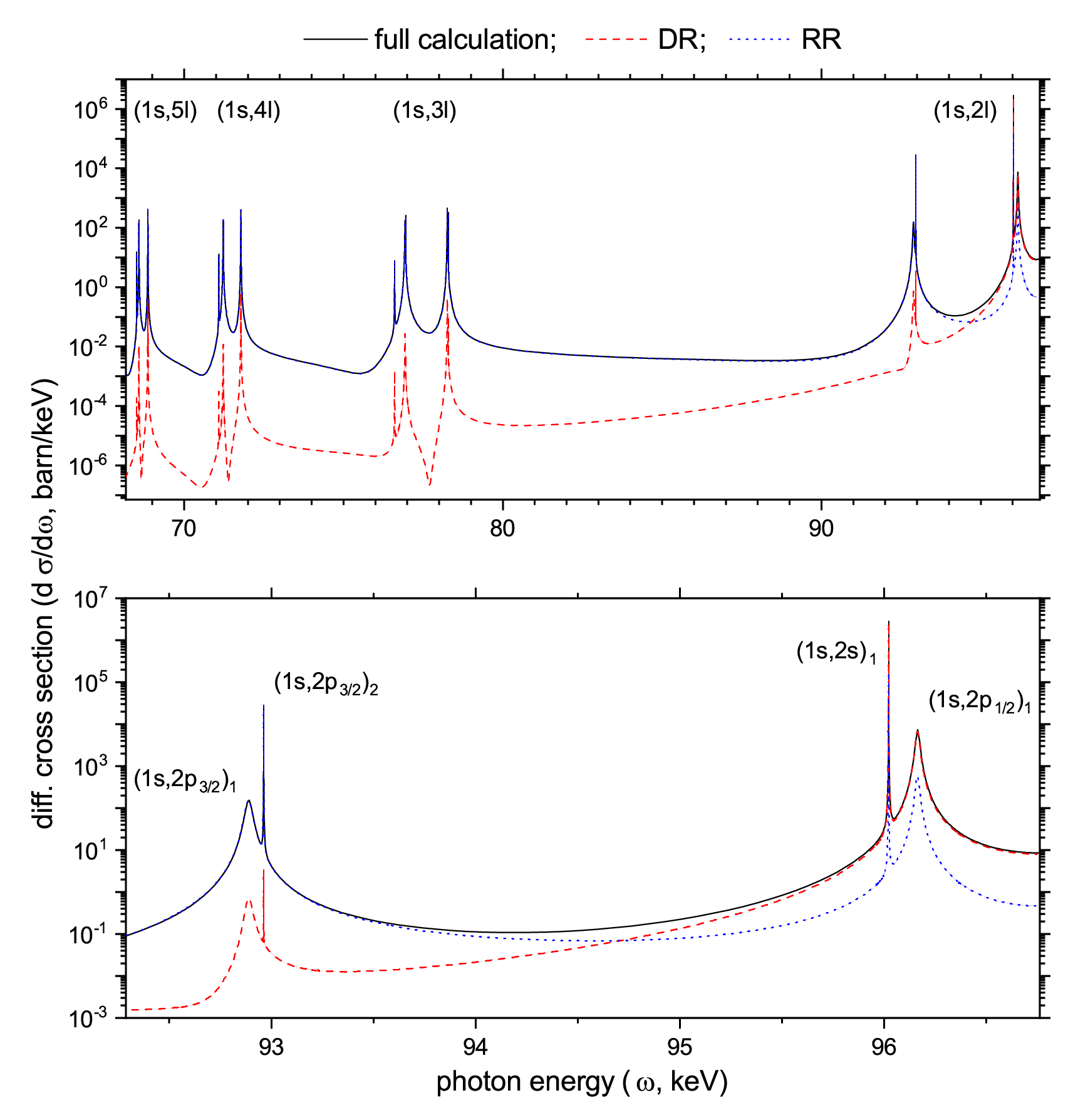}
\caption{Differential cross section (in barn/keV) as a function of the photon energy. The solid  (black) curve corresponds to the full calculation (same data as in Fig. \ref{fig1}). The dashed (red) curve represents the separated contribution of the DR channel. The dotted (blue) curve shows the separated contribution of the RR channel.}
\label{fig4}
\end{figure}

In this study, the initial energy of the system (the incident electron and the bound $1s$-electron) is chosen so that it is close to the energies of doubly excited $(2l,2l')$ states.
This condition can be written as
\begin{eqnarray}
E_{I}
&\approx&\label{eq2391}
E(2l,2l')
\,.
\end{eqnarray}
Therefore, the electron recombination occurs mainly through the DR channel, i.e., through the formation of the doubly excited states.
The fulfillment of the DR resonance condition \Eq{eq2391} is manifested in an increase in the total cross section (see Fig. \ref{fig00}).
In Fig. \ref{fig4} we present separately the contributions of the DR and RR channels to the differential cross section.
The RR channel is determined by disregard of the contributions of the doubly excited states in the full calculation. The amplitude of the DR channel is defined as the difference between the full amplitude and the amplitude of the RR channel. For the chosen energy of the incident electron, the interference between the DR and RR channels is generally very small, even in the photon energy regions where the DR and RR have similar values.

The most significant contribution to the DR channel is made by the singly excited $(1s,2l)$ states. This can be seen from the following: i) the DR channel proceeds through the formation of the $(2l,2l')$ states; ii) the most significant transitions are the $E1$ transitions in which only one of the electrons changes its quantum numbers. Accordingly, the main contribution to the DR channel is given by the following transitions
\begin{eqnarray}
\txt{U}^{91+}(1s) + e^{-}(\varepsilon)
&\to&\nonumber
\txt{U}^{90+}(2l,2l')
\\
&&\nonumber
\hspace{30pt}\downarrow
\\
&&\nonumber
\txt{U}^{90+}(1s,2l) + \gamma(E1)
\\
&&\nonumber
\hspace{30pt}\downarrow
\\
&&
\txt{U}^{90+}(1s)^2 + \gamma(E1)+\gamma'
\,.
\end{eqnarray}
In the RR channel, the initial state can effectively decay through any singly excited state. Thus, the DR channel is significantly suppressed in the photon energy region corresponding to the $(1s,3l)$, $(1s,4l)$ and $(1s,5l)$ cascade resonances.
We note that the contribution of the DR channel to the total cross section is dominant  (see \Fig{fig00}) due to the particularly large contribution of the resonance group $(1s,2l)$.
Fig. \ref{fig4} demonstrates that the Fano structure is more prominent in the DR channel than in the RR channel. 
We can expect that the Fano structure is more noticeable for lighter ions, where the DR channel is stronger relative to the RR channel.

\begin{figure}[ht]
\includegraphics[width=20pc]{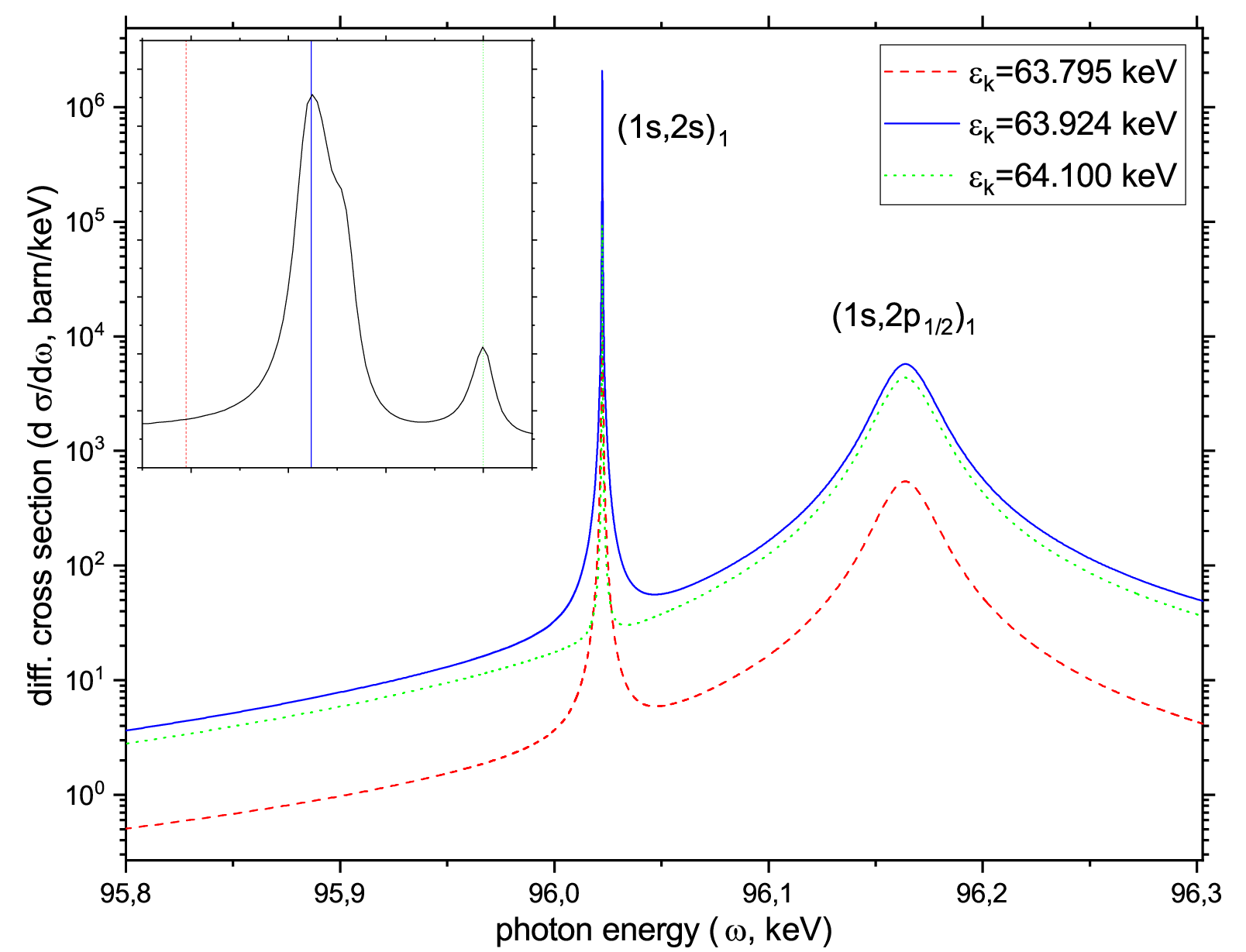}
\caption{Differential cross section as a function of the photon energy. The solid (blue), dotted (green) and dotted (red) curves correspond to the kinetic energy of the incident electron $\varepsilon_{\txt{k}}=63.924$ keV, $64.1$ keV and $63.795$ keV, respectively. The small inset in the upper-left corner shows the total cross section as a function of the incident-electron energy (see Fig.~\ref{fig00}), where the chosen electron energies are indicated by vertical lines of the corresponding color.}
\label{fig_id}
\end{figure}

It is worth considering in more detail the region of photon energy where the DR channel is not suppressed by the RR.
In order to differentiate the contributions of the RR and DR channels, we present the calculation results for several characteristic energies of the incident electron.
In particular, in Fig.~\ref{fig_id} the differential cross sections for the three selected energies are presented: i) the solid (blue) curve corresponds to the energy of the main resonance; ii) the dotted (green) curve corresponds to the energy of the smaller resonance corresponding to the $(2s)^2$ state; iii) the dotted (red) curve shows the differential cross section for energy far from resonances.
In the inset in the upper-left corner, we show the total cross section (see Fig.~\ref{fig00}) with vertical lines indicating the selected energies of the incident electron.
We can see that, in general, the difference in the energies of the incident electrons is manifested in an overall increase or decrease in the cross section.
The DR contribution increases the cross section by about 10 times.

\subsection{Importance of the Breit interaction}
\begin{figure}[ht]
\includegraphics[width=20pc]{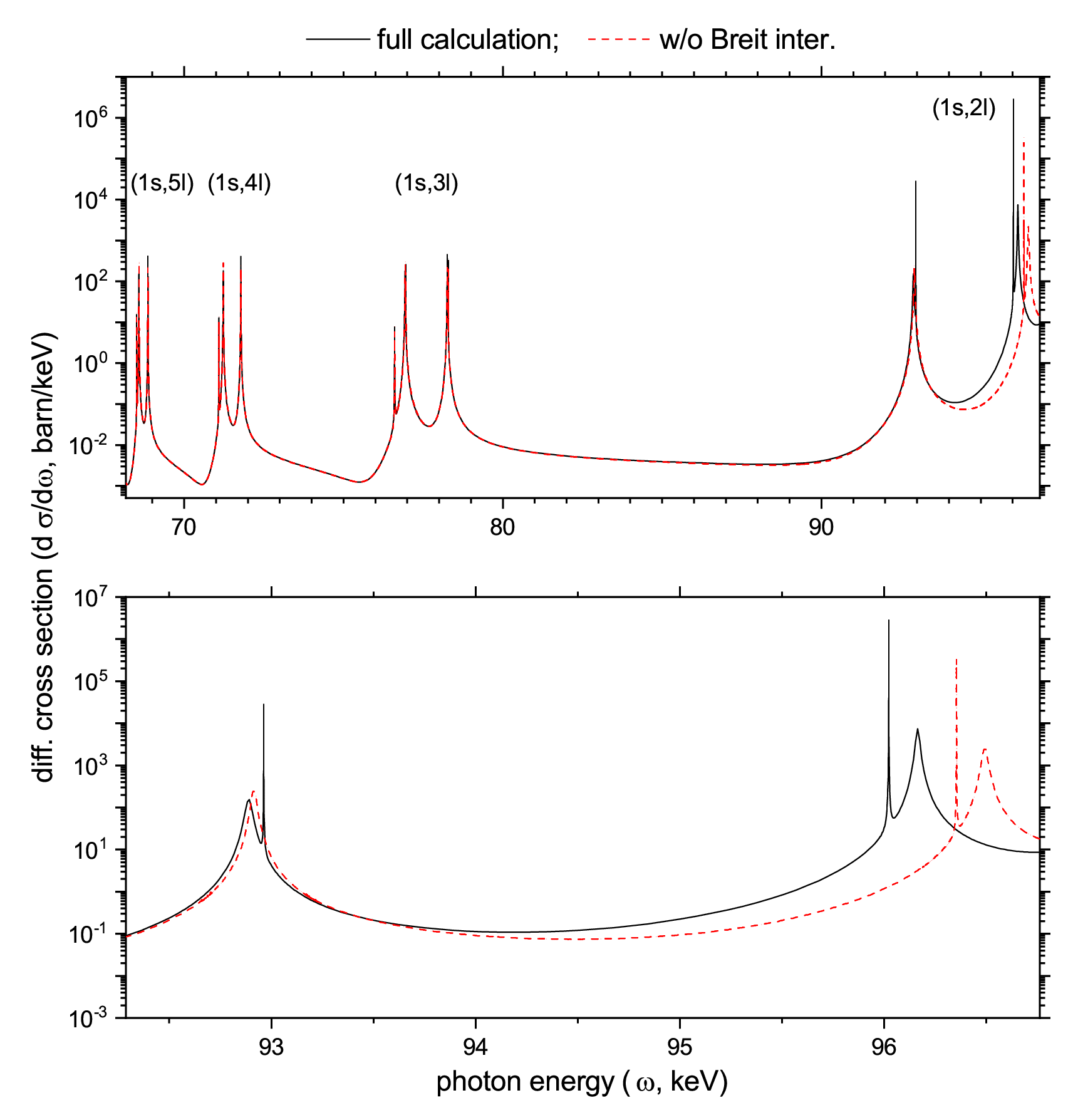}
\caption{Differential cross section (in barn/keV) as a function of the emitted photon energy. The solid (black) curve corresponds to the full calculation (the same data as in Fig. \ref{fig1}). The dashed (red) curve represents the result of the calculation without Breit interelectron interaction.}
\label{fig5}
\end{figure}

The interelectron interaction plays a crucial role for the DR channel. In particular, the contribution of the Breit interaction to the (differential) cross section is significant. In Fig. \ref{fig5} we show our results that demonstrate the role of the Breit interaction in the two-photon electron capture. We present a differential cross section calculated taking into account both the Coulomb and Breit (including the retardation \cite{lindgren95}; the so-called generalized Breit interaction) parts of the interelectron interaction, and also separately present the results of the calculation performed without the Breit interaction.

We can conclude that the Breit interaction affects the differential cross section in three different ways. 
First, it contributes to the energies of the singly and doubly excited states. This leads to the noticeable shift of both the positions of the cascade resonances in the differential cross section and the positions of the DR resonances in the total cross section.
Secondly, the Breit interaction determines the so-called Breit width of the energy levels, which is of great importance for some singly and doubly excited states \cite{lyashchenko2016}.
We see that disregard of the Breit interaction brings a significant change in the widths of some resonances, which can also cause some resonances becoming unnoticeable (e.g., the sharp resonance located near the energy of 93 keV in Fig.~\ref{fig5}).
Thirdly, it is the direct contribution of the Breit interaction to the rate of the formation of doubly excited states, since the formation occurs due to the interelectron interaction.
Disregard of the Breit interaction usually leads to a decrease in this rate, which reduces the DR resonances of the total cross section.
We see that the role of the Breit interaction is most significant for the cascade resonances corresponding to $(1s,2l)$ states, and decreases for other $(1s,nl)$ states as the principal quantum number $n$ increases. This is explained by the fact that the average orbital radius of the one-electron states increases with the $n$ (as $n^2$) and, consequently, the average interaction with the electric field of the atomic nucleus decreases. This makes electrons less relativistic and less sensitive to the QED effects, in particular, to the Breit interaction.

\subsection{Resonance approximation}\label{subsec_resapprox}
\begin{figure}[ht]
\includegraphics[width=20pc]{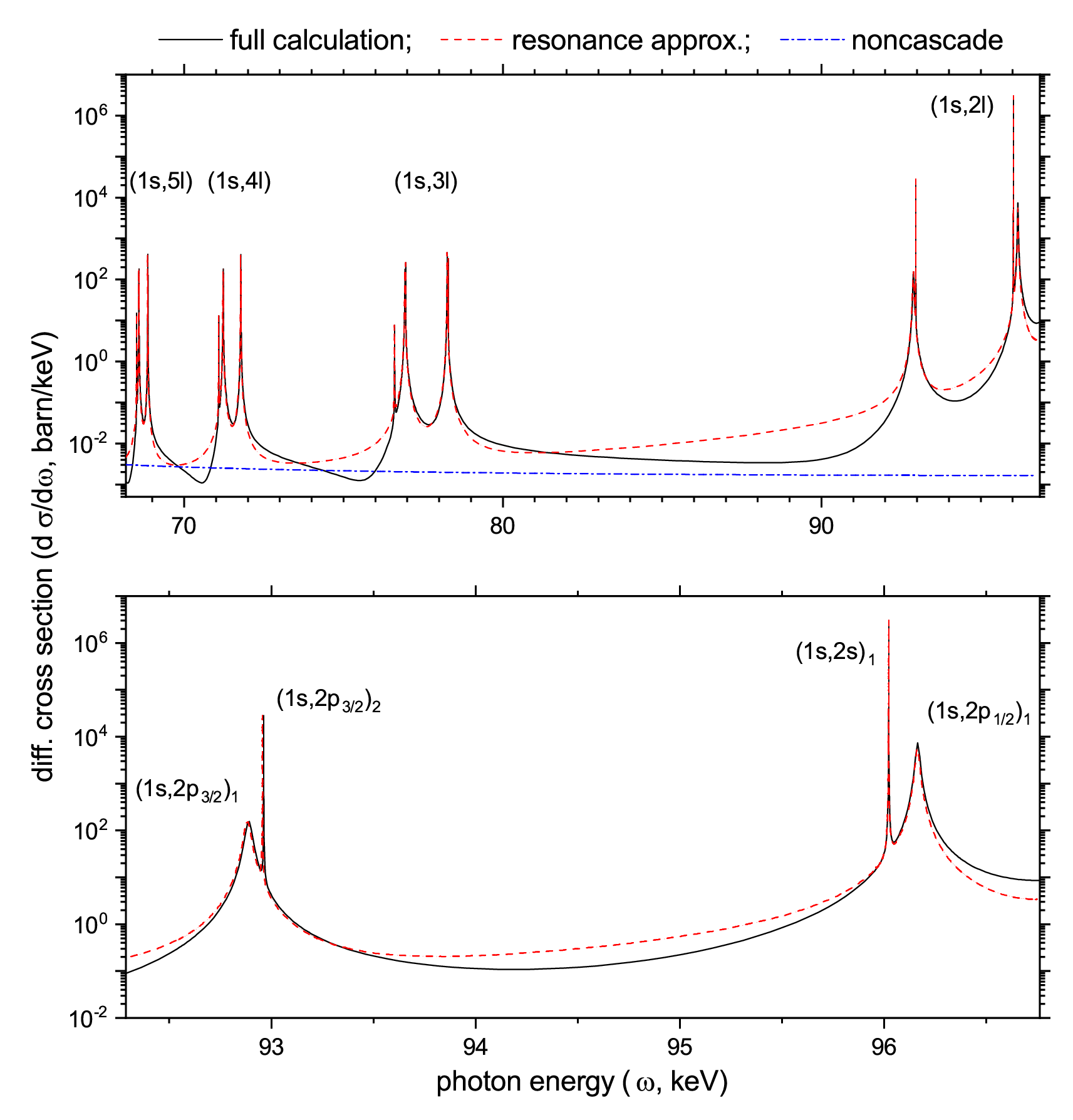}
\caption{Differential cross section (in barn/keV) as a function of the emitted photon energy. The solid (black) curve corresponds to the full calculation (the same data as in Fig. \ref{fig1}). The dashed (red) curve represents the result of calculation obtained within the resonance approximation. The dotted (blue) curve corresponds to the calculation where only the noncascade terms are taken into account.}
\label{fig6}
\end{figure}

The line profile corresponding to a single resonance can be effectively interpolated by the Lorentz profile.
Using this approach for a group of resonances leads to neglect of their interference \cite{Cowan_2023,kuhn2022PhysRevLett.129.245001}.
Here we introduce an appropriate approximation, which we call the resonance approximation.
We can see in Fig.~\ref{fig1} that the main contribution to the differential cross section is made by states corresponding to the cascade transitions.
In general, these are the singly excited $(1s,nl)$ and the doubly excited $(nl,n'l')$ states with the energies in the region $[E_{F},E_{I}]$. The resonant approximation includes: i) retaining only the cascade states in the summation over intermediate states in \Eq{3}; ii) neglecting the interference between the contributions of the retained states.
The implementation of the resonance approximation for the differential cross section yields 
\begin{eqnarray}
\frac{d^3\sigma^{\txt{(R)}}}{d\omega_1 d\Omega_1 d\Omega_2}
&=&\nonumber
\frac{\varepsilon}{p}
\frac{\omega_1^2 \omega_2^2}{(2\pi)^5}
\\
&\times&\nonumber
{\sum_{N}}'
\left(
\frac{
\left|\left(A^{*}_{k_1,\lambda_1}\right)_{FN} \right|^2
\left|\left(A^{*}_{k_2,\lambda_2}\right)_{NI} \right|^2
}
{(E_{F}+\omega_1-E_{N})^2+\frac{1}{4}\Gamma_{N}^2}
\right.
\\
&+&\label{dsection_lorentz}
\left.
\frac{
\left|\left(A^{*}_{k_2,\lambda_2}\right)_{FN} \right|^2
\left|\left(A^{*}_{k_1,\lambda_1}\right)_{NI} \right|^2
}
{(E_{I}-\omega_1-E_{N})^2+\frac{1}{4}\Gamma_{N}^2}
\right)
\,,
\end{eqnarray}
where the summation over the intermediate states in the amplitude \Eq{3} is replaced by the summation over the individual contributions of these states to the differential cross section.
The prime at the summation means that it runs only over the singly and doubly excited states with the energies within the region $[E_{F},E_{I}]$.
In Fig. \ref{fig6}, we compare the results obtained within the resonance approximation with the results of the full calculation.
The contribution of the noncascade terms are represented by a dashed dotted (blue) line.
Its relative contribution is quite small.
We can see that the resonance approximation is very accurate in the vicinity of every individual resonance.
It also gives quite reasonable results in the energy region of the same resonance group. However, this approximation becomes inapplicable when moving away from resonances, especially for the region between the groups of resonances.
Since the Fano structure is determined by the interference between the resonances, it can arise only beyond the resonance approximation.
In the energy region of the same resonant group, the resonant approximation can both increase and decrease the differential cross section, revealing the destructive and constructive role of interference.
In the energy region between resonance groups, the resonant approximation significantly increases the cross section, revealing the destructive role of the interference between the resonances.

\subsection{Single-photon approximation}\label{single-photon_approximation}
In the experiment on the DR with H-like uranium ion \cite{bernhardt11}, the electron capture was registered by  charge change of the ions
\begin{eqnarray}
U^{91+}(1s) + e(\varepsilon)
&\to&\nonumber
U^{90+}(2l,2l')\\
&&\nonumber
\hspace{30pt}\downarrow\\
&&
U^{90+}(1s)^2 + \gamma +\gamma' + \cdot\cdot\cdot
\end{eqnarray}
So, the emitted photons were not detected in this experiment.
Only the number of ions with changed charge was measured.
The main contribution to this process was made by the electron capture to the singly excited states.
The major theoretical studies considered only a part of this process
\begin{eqnarray}
U^{91+}(1s) + e(\varepsilon)
&\to&\nonumber
U^{90+}(2l,2l')\\
&&\nonumber
\hspace{30pt}\downarrow\\
&&
U^{90+}(1s,2l) + \gamma
\,,
\end{eqnarray}
where the emission of the second and other photons was neglected.
The calculated cross section for the one-photon transitions was used for description of the experimental data.
The obtained theoretical and experimental results are in good agreement \cite{bernhardt11}.
The absence of further transitions from the singly excited states to the ground state determines the single-photon approximation.
This approximation was first considered in \cite{zakowicz04}, and calculations of some properties of this process beyond the single-photon approximation were presented within the quasirelativistic approach.

In this subsection, we investigate the single-photon approximation.
In order to apply it for the two-photon electron capture, we do the following: i) use the resonance approximation introduced in Subsection \ref{subsec_resapprox}; ii) perform integration over the energy of one of the emitted photons in the expression for the differential cross section \Eq{dsection_lorentz}.
Within the single-photon approximation the differential cross section for the two-photon electron capture reads as
\begin{eqnarray}
\frac{d\sigma_{I\to F}}{d\Omega_2}
&=&
\sum_{N}
\frac{d\sigma_{I\to N}^{(\txt{1\gamma\,approx})}}{d\Omega_2}
\frac{\Gamma_{N}^{(1\gamma,N\to F)}}{\Gamma_{N}}
\,.
\end{eqnarray}
The detailed derivation is presented in Appendix~\ref{appendix2}.
The differential cross section $\frac{ d\sigma_{I\to N}^{(\txt{1\gamma\,approx})}}{d\Omega_2}$ is defined by \Eq{eq67129087}.
Its physical meaning is the cross section of the electron recombination into the state $N$ with the one-photon emission.
The width $\Gamma_{N}$ is the total width of the state $N$,
$\Gamma_{N}^{(1\gamma,N\to F)}$ denotes the partial one-photon width corresponding to the transition of the singly excited state $N$ to the final state $F$.

The total width of the singly excited states ($\Gamma_{N}$) is determined by the radiation width. In the case of the singly excited states with the total angular momentum not equal to zero ($J\ne0$), the total widths are determined by the one-photon widths $\Gamma_{N}\approx\Gamma_{N}^{(1\gamma)}$.
For the $(1s,2l)$ states, the one-photon widths are determined by one-photon transitions to the ground state $\Gamma_{N}^{(1\gamma)} \approx \Gamma_{N}^{(1\gamma,N\to F)}$.
Therefore, we can substitute unit for the fraction $\frac{\Gamma_{N}^{(1\gamma,N\to F)}}{\Gamma_{N}}$.
For the $(1s,nl)$ states with $n\ge3$, we also substitute unit for this fraction within the one-photon approximation. However, for the $(1s,nl)$ states with $n\ge3$, this may lead to a loss of accuracy.
Since we are considering the two-photon electron capture, the intermediate singly excited states ($N$) with zero total angular momentum do not contribute, because they cannot decay to the ground state by emitting a single photon.
Accordingly, the cross section of the two-photon electron capture within the single-photon approximation can be written as
\begin{eqnarray}
\frac{d\sigma_{I\to F}}{d\Omega_2}
&\approx&\label{17}
\sum_{N (J\ne0)}
\frac{ d\sigma_{I\to N}^{(\txt{1\gamma\,approx})}}{d\Omega_2}
\,,
\end{eqnarray}
where the summation runs over the singly excited states with nonzero total angular momentum.

It is useful to consider the capture of an electron with the emission of two and more photons.
In this case, by integrating the energies of all the emitted photons, integrating over the angular variable and summing the polarizations of all the emitted photons, with the exception of the angular variables and polarization of the resonant photon, which corresponds to the transition from $I\to N$, we come to the expression
\begin{eqnarray}
\frac{d\sigma_{I\to F}}{d\Omega_2}
&=&\nonumber
\sum_{N}
\frac{ d\sigma_{I\to N}^{(\txt{1\gamma\,approx})}}{d\Omega_2}
\frac{\Gamma_{N}^{(N\to F)}}{\Gamma_{N}}\\
&\approx&\label{18}
\sum_{N}
\frac{ d\sigma_{I\to N}^{(\txt{1\gamma\,approx})}}{d\Omega_2}
\,.
\end{eqnarray}
In this case, since many photon transitions have been taken into account,
$\Gamma_{N}^{(N\to F)}\approx\Gamma_{N}$ becomes a good approximation for all the singly excited states.

So far, we have not considered the electron capture to the ground state with emission of a single photon.
However, this makes a noticeable contribution to the total electron capture cross section.
In this process, we can also distinguish two channels: the RR and DR.
The RR channel is nonresonant and gives a correction to the cross section, which varies slightly depending on the incident energy \cite{eichler1995PhysRevA.51.3027}.
The DR channel is resonant, but it proceeds through the TEOP transition, which is usually negligible.
Accordingly, the one-photon electron capture can be easily taken into account with high accuracy.

To obtain the total cross of the electron capture with the emission of two or more photons in the single-photon approximation, we
perform integration over the angles of the emitted photon ($\Omega_2$) in Eq.(\ref{18}), summation and averaging over the polarizations of the electrons and photon.
This cross section together with the contribution of the one-photon electron capture, gives the cross section measured in the experiment \cite{bernhardt11}, where only the charge change of the ions was recorded.

In Fig.~\ref{fig00}, the dashed (red) curve represents the total cross section of the electron capture calculated within the single-photon approximation.
We note that, since we are studying the two-photon electron capture, only the singly excited states with nonzero total angular momenta ($J\ne0$) are taken into account as final states in calculations within this approximation.
The cross section, which takes into account all the final states, is presented in
\cite{lyashchenko2015}.
This cross section can be compared with the results of the full calculation given by the solid (black) curve.
The single-photon approximation enlarges the cross section by about 5\% in the nonresonance region and by about 1\% in the resonance region.

The single-photon approximation can be applied to two- and more electron systems for description of experiments on the radiative electron capture in which the emitted photons are not recorded \cite{Nakamura2008PhysRevLett.100.073203,bernhardt11,Huang_2018}.

\section{Summary}
We have studied the two-photon capture of an electron with the energy corresponding to the strongest DR resonances.
For the first time, we conducted a detailed study within the QED theory of the photon emission spectrum taking into account four main groups of the cascade resonances ($(1s, nl)$, where $n=2-5$).
Special attention was paid to the region of low energy photons, where the infrared divergence requires a special approach.
The influence of the DR resonances on the emission spectrum was shown.
We demonstrated that the contribution of the Breit interaction is very important.
The Breit interaction makes a significant contribution to the positions and widths of many resonances and can qualitatively change the emission spectrum.
We analyzed the widely used resonant approximation, studying its accuracy.
We have shown that this approximation gives suitable results for the photon energies near the cascade resonances.
However, it may not work in the regions between close resonances if there is strong interference between them.
Moreover, it failed in the region between the resonance groups.
Finally, we demonstrated how the two-photon electron capture is related to an experimental setup where the DR was investigated by measuring the number of ions that captured an electron (without registering the emitted photons).
The investigation of the photon emission spectrum in the process of two-photon electron capture provides an opportunity to test the QED theory by studying the atomic structure and dynamics in strong fields.

\begin{acknowledgments}
The work is supported by the National Key Research and Development Program of China under Grant No. 2022YFA1602501 and the National Natural Science Foundation of China under Grant No. 12011530060.
The work of O.Y.A., K.N.L. on sections III E and F was supported solely by the Russian Science Foundation under Grant No. 22-12-00043.
K.N.L. and O.Y.A. was supported by the Chinese Academy of Sciences (CAS) Presidents International Fellowship Initiative (PIFI) under Grant Nos. 2022VMC0002 and 2018VMB0016, respectively. 
\end{acknowledgments}

\appendix
\section{Single-photon approximation}
\label{appendix2}
We present the application of the single-photon approximation to the cross section of the two-photon electron capture. First, we consider the cross section within the resonance approximation Eq.(\ref{dsection_lorentz}). Secondly, we perform integration over the energy of the photon $\omega_1$ 
\begin{eqnarray}
\frac{d^2\sigma_{I\to F}}{d\Omega_1 d\Omega_2}
&\approx&\nonumber
\int_0^{\omega_{\txt{max}}/2} 
d\omega_1 \,
\frac{\varepsilon}{p}
\frac{\omega_1^2 \omega_2^2}{(2\pi)^5}
\\
&&\nonumber
\hspace{-40pt}
\times
\sum_{N}
\left(
\frac{
\left|\left(A^{*}_{k_1,\lambda_1}\right)_{FN} \right|^2
\left|\left(A^{*}_{k_2,\lambda_2}\right)_{NI} \right|^2
}
{(E_{F}+\omega_1-E_{N})^2+\frac{1}{4}\Gamma_{N}^2}
\right.\\
&&\label{19}
\hspace{-40pt}
\left.
+
\frac{
\left|\left(A^{*}_{k_2,\lambda_2}\right)_{FN} \right|^2
\left|\left(A^{*}_{k_1,\lambda_1}\right)_{NI} \right|^2
}
{(E_{I}-\omega_1-E_{N})^2+\frac{1}{4}\Gamma_{N}^2}
\right)
.
\end{eqnarray}
Only one of the terms in large brackets is resonant within the given integration region. 
In our case, the first term is resonant, while the second term is not resonant and can be omitted.
\begin{eqnarray}
\frac{d^2\sigma_{I\to F}}{d\Omega_1 d\Omega_2}
&\approx&\nonumber
\int_0^{\omega_{\txt{max}}/2} 
d\omega_1 \,
\frac{\varepsilon}{p}
\frac{\omega_1^2 \omega_2^2}{(2\pi)^5}
\\
&&\label{20}
\hspace{-20pt}
\times
\sum_{N}
\frac{
\left|\left(A^{*}_{k_1,\lambda_1}\right)_{FN} \right|^2
\left|\left(A^{*}_{k_2,\lambda_2}\right)_{NI} \right|^2
}
{(E_{F}+\omega_1-E_{N})^2+\frac{1}{4}\Gamma_{N}^2}
\,.
\end{eqnarray}
The main contribution to the integral is given by the resonant region around $\omega_1 \approx E_{N}-E_{F}$. Within the resonant region the matrix elements $\left(A^{*}_{k_1,\lambda_1}\right)_{FN}$ and $\left(A^{*}_{k_2,\lambda_2}\right)_{NI}$ as well as the factors $\omega_1^2$ and $\omega_2^2$ are slowly changing function of $\omega_1$ and can be substituted by their values in the resonance
\begin{eqnarray}
\frac{d^2\sigma_{I\to F}}{d\Omega_1 d\Omega_2}
&\approx&\nonumber
\sum_{N}
\frac{\varepsilon}{p}
\frac{(\omega_{1}^\txt{res})^2(\omega_2^\txt{res})^2}{(2\pi)^5}
\\
&\times&\nonumber
\left|
\left(A^{*}_{k_1^\txt{res},\lambda_1}\right)_{FN} 
\right|^2
\left|
\left(A^{*}_{k_2^\txt{res},\lambda_2}\right)_{NI} 
\right|^2\\
&\times&\label{13}
\int_0^{\omega_{\txt{max}}/2}
\!\!
\frac{d\omega_1}{(E_{F}+\omega_1-E_{N})^2+\frac{1}{4}\Gamma_{N}^2}
,
\end{eqnarray}
where $k_1^\txt{res}=(E_{N}-E_{F},{\bf{k}}_1)$, $k_2^\txt{res}=(E_{I}-E_{N},{\bf{k}}_2)$.
We can utilize the approximation
\begin{eqnarray}
&&\nonumber
\int_0^{\omega_{\txt{max}}/2}
\frac{d\omega_1}{(E_{F}+\omega_1-E_{N})^2+\frac{1}{4}\Gamma_{N}^2}
\\
&\approx&\label{12}
\int_{-\infty}^{\infty}
\frac{d\omega_1}{(E_{F}+\omega_1-E_{N})^2+\frac{1}{4}\Gamma_{N}^2}
=
\frac{2\pi}{\Gamma_{N}}
\,.
\end{eqnarray}
Using \Eq{12} for \Eq{13}, performing the integration over $\Omega_1$, the summation over the photon polarization  ($\lambda_1$), and the summation over the projection of the final state $M_{F}$, we obtain the following expression
\begin{eqnarray}
\frac{d\sigma_{I\to F}}{d\Omega_2}
&\approx&\nonumber
\sum_{N}
\left\{
2\pi \frac{\varepsilon}{p}
\frac{(\omega_2^\txt{res})^2}{(2\pi)^3}
\left|
\left(A^{*}_{k_2^\txt{res},\lambda_2}\right)_{NI} 
\right|^2
\right\}
\frac{1}{\Gamma_{N}}\\
&&\label{14}
\hspace{-40pt}
\times
\left[
\sum_{M_{F},\lambda_1}
\frac{(\omega_{1}^\txt{res})^2}{(2\pi)^2}
\int d\Omega_1
\left|
\left(A^{*}_{k_1^\txt{res},\lambda_1}\right)_{FN} 
\right|^2 
\right]
\,.
\end{eqnarray}
The term in the square brackets does not depend on $M_{N}$ and is equal to the one-photon partial width of the state $N$ 
\begin{eqnarray}
\hspace{-20pt}
\sum_{M_{F},\lambda_1}
\!\!
\frac{(\omega_{1}^\txt{res})^2}{(2\pi)^2}
\!
\int d\Omega_1
\left|
\left(A^{*}_{k_1^\txt{res},\lambda_1}\right)_{FN} 
\right|^2 
\!&=&\!
\Gamma_{N}^{(1\gamma,N\to F)}
\,.
\end{eqnarray}
The term in curly brackets together with the summation over the projection $M_{N}$ gives the differential cross section within the one-photon approximation
\begin{eqnarray}
\frac{ d\sigma_{I\to N}^{(\txt{1\gamma\,approx})}}{d\Omega_2}
&=&\label{eq67129087}
2\pi \frac{\varepsilon}{p}
\frac{(\omega_2^\txt{res})^2}{(2\pi)^3}
\left|
\left(A^{*}_{k_2^\txt{res},\lambda_2}\right)_{NI} 
\right|^2
\,.
\end{eqnarray}
Accordingly, we can write
\begin{eqnarray}
\frac{d\sigma_{I\to F}}{d\Omega_2}
&\approx&
\sum_{N}
\frac{ d\sigma_{I\to N}^{(\txt{1\gamma\,approx})}}{d\Omega_2}
\frac{\Gamma_{N}^{(1\gamma,N\to F)}}{\Gamma_{N}}
\,.
\end{eqnarray}
In general, the full width of any state can be represented as sum of the radiative and Auger widths: $\Gamma_{N}=\Gamma_{N}^{(\txt{rad})}+\Gamma_{N}^{(\txt{Auger})}$.
If we assume that the main contribution is given by the singly excited states $(1s,2l)$ with the total angular momentum not equal to zero ($J\ne 0$),
we get $\Gamma_{N}^{(1\gamma,N\to F)}\approx \Gamma_{N}$.
Finally, we get the following relation between the differential cross section of the two-photon electron capture and the differential cross section within the single-photon approximation
\begin{eqnarray}
\frac{d\sigma_{I\to F}}{d\Omega_2}
&\approx&
\sum_{N}
\frac{ d\sigma_{I\to N}^{(\txt{1\gamma\,approx})}}{d\Omega_2}
\,.
\end{eqnarray}
By integrating over the angular variable $\Omega_2$ and summing and averaging over the polarizations of the electrons and photons, we obtain the total cross section as
\begin{eqnarray}
\sigma_{I\to F}
&\approx&\nonumber
\sum_{\mu,m_b, M_{I} \lambda_2}\sum_{N}
\int d\Omega_2 \, \frac{ d\sigma_{I\to N}^{(\txt{1\gamma\,approx})}}{d\Omega_2}
\\
&\times&
\frac{1}{2(2j_b+1)}
\,,
\end{eqnarray}
where $j_b$ is the total angular momentum of the bound electron in the initial state.

%

\end{document}